\documentclass[article,twocolumn]{revtex4}
\usepackage{graphicx}
\usepackage{amsmath,amssymb}
\usepackage{epsfig}
\usepackage{amsfonts, color, soul}

\newcommand \beq {\begin{equation}}
\newcommand \bea {\begin{eqnarray} \nonumber }
\newcommand \eeq {\end{equation}}
\newcommand \eea {\end{eqnarray}}
\newcommand{\beqa}{\begin{eqnarray}}
\newcommand{\eeqa}{\end{eqnarray}}

\newcommand{\Y}{${\rm CheY}$}
\newcommand{\Yp}{${\rm CheY_p}$}
\newcommand{\Ym}{${\rm CheY^*}$}

\begin{document}

\title{The switching dynamics of the bacterial flagellar motor}
\author{Siebe B. van Albada}
\affiliation{FOM Institute for Atomic and Molecular Physics,
Kruislaan 407, 1098 SJ Amsterdam, The Netherlands.}

\author{Sorin T\u{a}nase-Nicola}
\affiliation{FOM Institute for Atomic and Molecular Physics,
Kruislaan 407, 1098 SJ Amsterdam, The Netherlands.}

\author{Pieter Rein ten Wolde}
\affiliation{FOM Institute for Atomic and Molecular Physics,
Kruislaan 407, 1098 SJ Amsterdam, The Netherlands.}

\begin{abstract}
  Many swimming bacteria are propelled by flagellar motors that
  stochastically switch between the clockwise and counterclockwise
  rotation direction. While the switching dynamics are one of the most
  important characteristics of flagellar motors, the mechanisms that
  control switching are poorly understood.  We present a
  statistical-mechanical model of the
  flagellar rotary motor, which consist of a number of stator proteins
  that drive the rotation of a ring of rotor proteins, which in turn
  drives the rotation of a flagellar filament.  At the heart of our
  model is the
  assumption that the rotor protein complex can exist in two
  conformational states corresponding to the two respective rotation
  directions, and that switching between these states depends on
  interactions with the stator proteins. This naturally couples the
  switching dynamics to the rotation dynamics, making the switch
  sensitive to torque and speed. Another key element of our model is
  that after a switching event, it takes time for the load to build
  up, due to polymorphic transitions of the filament. Our model
  predicts that this slow relaxation dynamics of the filament, in
  combination with the load dependence of the switching frequency,
  leads to a characteristic switching time, in agreement with recent observations.
\end{abstract}

\maketitle

\newpage

\section{Introduction}
The bacterium {\em Escherichia coli} can swim toward attractants and
away from various noxious chemicals. It is
propelled by several flagella. Each flagellum is under the action of a
rotary motor, which can rotate either in a clockwise (CW) or a
counterclockwise (CCW) direction. When all the motors run in the
counterclockwise direction, the flagella form a helical bundle and the
bacterium swims smoothly. When one motor switches direction to run in
the clockwise direction, however, the connected flagellar filament
disentangles from the bundle, and the bacterium performs a so-called
tumble. These tumble events randomize the cell's trajectory, and it is
the modulation of their occurrence that allows {\em E. coli} to
chemotax.  Here, we present a statistical-mechanical model that describes
the switching dynamics of the rotary motor.

\begin{figure}[t]
\includegraphics[width=8cm]{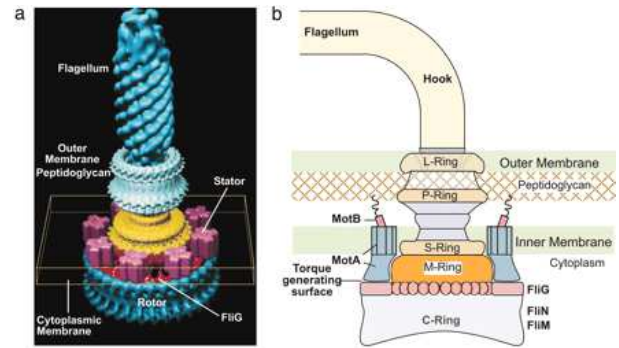}
\caption{Cartoon of the flagellar motor of the bacterium {\em
    E. coli.} The figure is courtesy of DeRosier \cite{Thomas06}.
\label{fig:cartoon}}
\end{figure}

A cartoon of the bacterial flagellar motor is shown in
Fig. \ref{fig:cartoon} \cite{Thomas06}. It consists of a protein
complex called the rotor, and a number of stator proteins that are
fixed in the inner membrane and the peptidoglycan layer. Interactions
between the stator proteins and a ring of FliG proteins of the rotor
protein complex drive the rotation of the rotor, and thereby the
rotation of the flagellum, which is connected to the rotor.
 The
rotation direction is determined by the concentration
of the phosphorylated form of the messenger protein \Y, which binds
to the ring of FliM proteins of the rotor protein complex.

The bacterial flagellar motor has a number of characteristics
that reflect its design principles. One is the
torque-speed relationship of the motor, which has recently been
modeled by Oster and coworkers \cite{Xing:2006dq}. Another
 is the power spectrum of the motor switching dynamics, which
 reflects the time scales on which the motor switches direction.
 Recently, Cluzel and coworkers measured power spectra for mutant
 cells by monitoring the rotation of a $0.5~\mu{\rm m}$ latex bead
 connected to a
 flagellum. \cite{Korobkova04,Korobkova06}. In these cells, 
 a mutant CheY protein, \Ym, was stably
 preexpressed \cite{Korobkova04,Korobkova06}. This protein mimics the
 effect of ${\rm CheY_p}$, but its
 concentration is not affected by the dynamics of the
 chemotaxis signaling network.
The power spectra of these mutant cells therefore reflect the
intrinsic switching dynamics of the motor.

Intriguingly, the power spectra of these \Ym\ mutant cells are not
consistent with a two-state Poisson process, in which the switching
events are independent and the CW and CCW intervals are uncorrelated
and exponentially distributed \cite{Korobkova06}.  They exhibit a
distinct peak
at around $1 \,{\rm s}^{-1}$ \cite{Korobkova06}, which means that
there is a characteristic frequency at which the motor switches. 
 Cluzel
and coworkers \cite{Korobkova06} suggest that an earlier model
developed by Duke and Bray \cite{Duke01} might be able to explain the
power spectrum.
This model, however, is based on an Ising system, which is a
mesoscopic equilibrium system. Such a system cannot exhibit a peak
in the power spectrum \cite{VanKampen92,ThesisVanAlbada,Tu08}. The peak
means that the switching dynamics is coupled to a non-equilibrium
process.

We argue that to explain the switching dynamics of the motor, we have
to integrate a description of the switching dynamics of the rotor with
a description of both the flagellum dynamics and the dynamics of the
stator proteins that drive the rotation of the rotor.  The rotor
protein complex is modeled as an Monod-Wyman-Changeux (MWC) model \cite{Monod65}, in which
the proteins of the complex collectively switch between a clockwise
and a counterclockwise conformational state. Interactions between the
stator proteins and the FliG proteins of the rotor do not only drive
the rotation of the rotor, but also continually change the relative
stability of the two conformational states of the rotor. This couples
switching to the non-equilibrium process of rotation, and makes the
switch sensitive to torque and speed. Our model predicts that the
probability for the rotor to switch increases with the load in the
zero-load regime, as observed experimentally \cite{Fahrner03}. But, to
fully describe the switching of the motor, the rotor's switching and
rotation dynamics have to be coupled to the conformational dynamics of
the flagellum.

Bacterial flagella can exist in different polymorphic states
\cite{Calladine:1975gl,Macnab:1977wm,Hotani:1982nq,Darnton:2007hc},
which are either left-handed or right-handed helices.  When the motor
runs in the CCW direction, the flagellum adopts a left-handed, normal
state \cite{Turner00}, while if it runs in the CW direction, it adopts
a right-handed, semi-coiled or curly state \cite{Turner00}. By pulling
on a single flagellum using optical tweezers, Darnton and Berg
recently observed that transitions between these polymorphic forms
occur in discrete steps, during which elastic strain energy is
released \cite{Darnton:2007hc}. We argue that the change in the torque
upon a motor reversal induces a polymorphic transition that proceeds
via a similar series of discrete steps.  Since in each of these steps
strain energy is released, the torque on the motor, and hence the
switching propensity, remains low. However, when the flagellum
ultimately reaches its final polymorphic form, the strain energy can
no longer be released and the torque on the motor increases. Due to
the exponential dependence of the motor's switching propensity on the
applied torque, the switching propensity now increases rapidly, giving
rise to the characteristic switching time \cite{Korobkova06}.

\section{The stator-rotor interaction}
The interaction between the stator proteins and the rotor proteins is
modeled according to the model of Oster and coworkers \cite{Xing:2006dq},
which is based on the description of Blair and coworkers
\cite{Kojima01}. Since our model builds upon this model, we briefly
describe its main ingredients. For details, we refer to the {\em
  Supporting Information}.

\begin{figure}[t]
{\bf \sf A}\includegraphics[width=4cm]{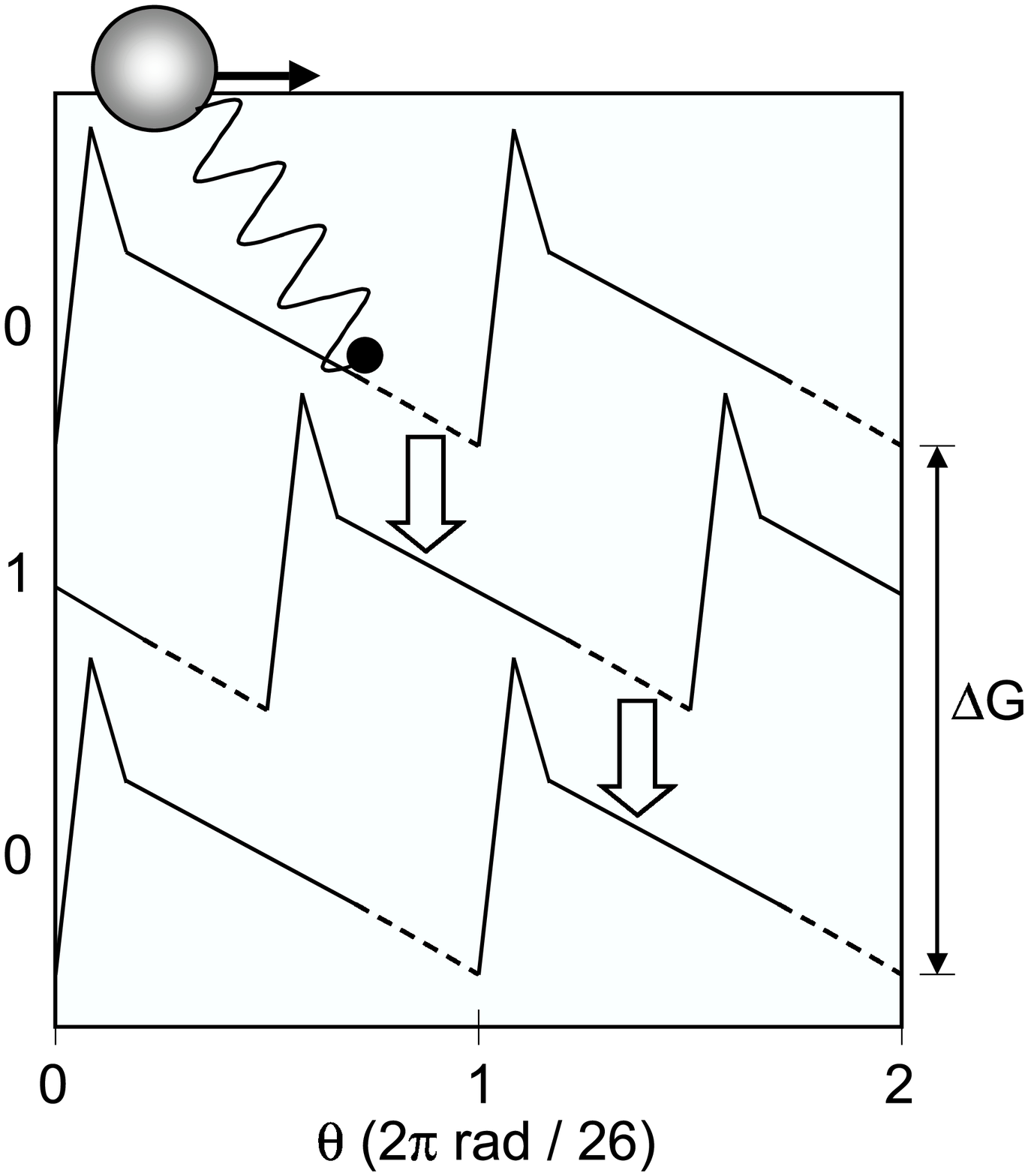}
{\bf \sf B}\includegraphics[width=4cm]{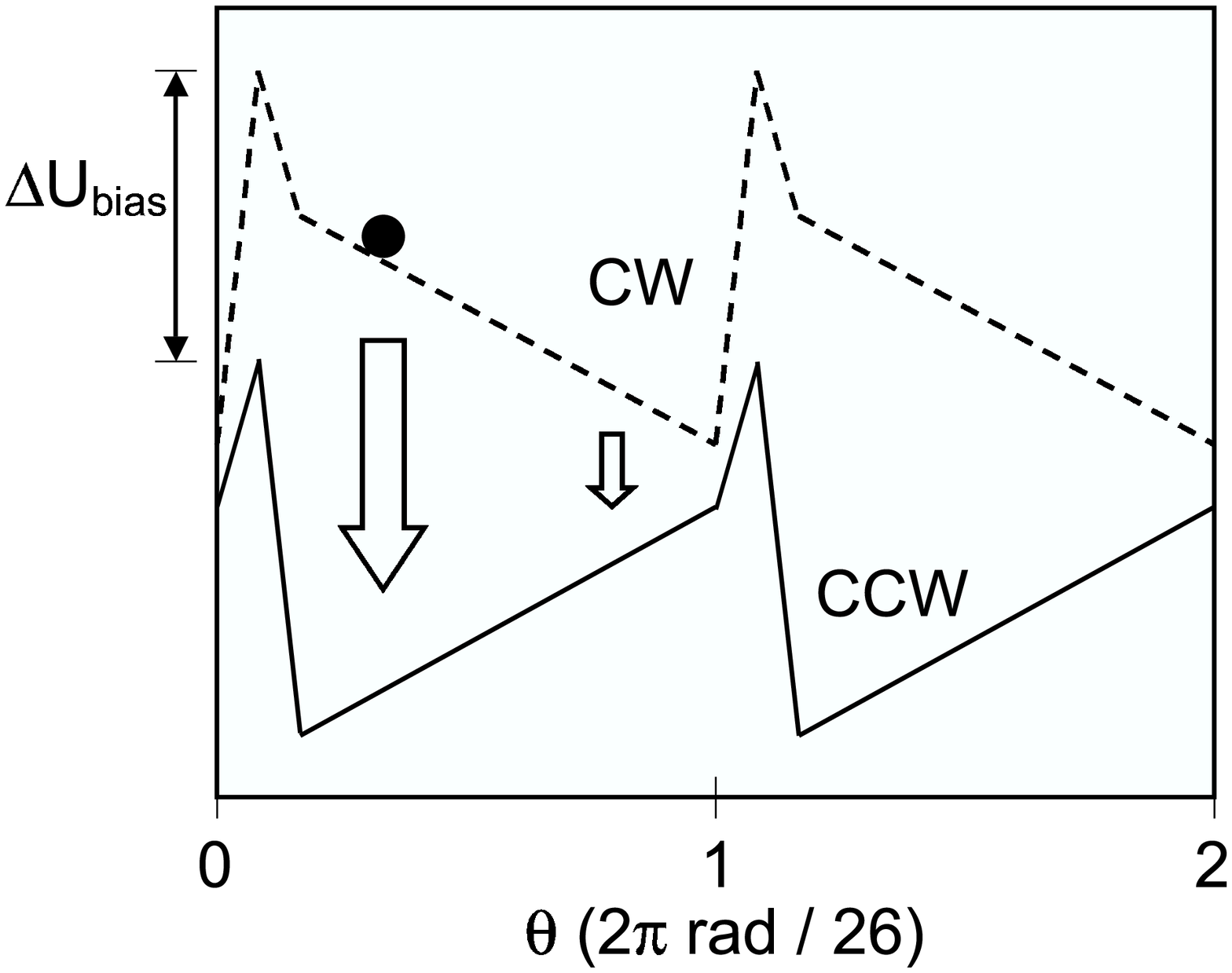}
\caption{\label{fig:OsterModel} Energy surfaces for the interaction
  between the rotor complex and one stator protein. {\bf \sf A}: Energy
  surfaces corresponding to the two conformational states of the stator
  for a given conformational (CW) state of the
  rotor protein complex, according to
  \cite{Kojima01,Xing:2006dq}. The dashed lines denote the hopping windows. The thermodynamic driving force is $\Delta G = 2 e \times
  {\rm pmf}$. {\bf \sf B}: Two energy surfaces of the motor,
  corresponding to the CCW and CW states of the rotor, for a given
  conformational state of the stator protein; the two surfaces are
  assumed to be each other's mirror image. In total, each rotor-stator
  interaction is characterized by 4 surfaces, corresponding to the
  $2\times2$ conformational states of the stator and rotor proteins.}
\end{figure}

According to the proposal of Blair and Oster, the motor cycle of each stator
protein consists of two ``half strokes''.
During the first power stroke of a stator protein, two protons bind
MotB residues of the stator protein \cite{Kojima01}
(Fig. \ref{fig:cartoon}). This leads to a thermally activated
conformational transition, which allows one MotA loop to exert a force
on one FliG protein of the rotor protein complex
(Fig. \ref{fig:cartoon}). During the second
stroke, the recovery stroke, the two protons are released to the
cytoplasm, triggering another conformational transition of the
stator, and allowing another MotA loop to exert a force on the
FliG protein.
At the end of this cycle, the rotor has
advanced by an angle of $2
\pi$ divided by 26, the number of FliG proteins within the
ring. The force exerted by MotA is modeled as a constant force along
an energy surface, and the conformational transitions are described as {\em hops}
between the two respective surfaces (Fig. \ref{fig:OsterModel}).

The rotation dynamics of the rotor is given by  the
following overdamped Langevin equation:
\begin{equation}
\label{eq:Lan}
\gamma_{\rm R} \frac{d\theta_{\rm R}}{dt} = -\sum_{j=1}^{N_{\rm S}}\frac{\partial
  U_{s_j}(\theta_j)}{\partial \theta_{\rm R}} +
  F_{\rm L} + \eta_{\rm R} (t)
\end{equation}
Here, $\gamma_{\rm R}$ is the friction coefficient of the rotor;
$U_{s_j} (\theta_j)$ is the rotor-stator interaction energy as a
function of $\theta_j\equiv \theta_{\rm R}-\theta_{{\rm S}_j}$ (Fig. \ref{fig:OsterModel}), where $\theta_{\rm R}$ denotes the rotor rotation
angle, $\theta_{{\rm S}_j}$ the angle of the immobile stator
protein $j$, and $s_j$ is a binary variable denoting the conformational
state of stator $j$; $\eta_{\rm R}(t)$ is a Gaussian white noise term
of magnitude $\sqrt{2k_{\rm B}T \gamma_{\rm R}}$; $N_{\rm S}$ is the
number of stator proteins. The torque $F_{\rm L}$ denotes the external
load. As discussed in \cite{Elston00,Elston00_2,Xing:2006dq}, for the
system studied here, the torque-speed curves under conservative load
and viscous load are identical. However, as we will show below, the
type of load does markedly affect the
switching dynamics.

The {\em hopping} rate for a stator protein to go from
one energy surface to another depends upon the free-energy barrier
separating them. We make the
phenomenological assumption that the hopping rate depends
exponentially on the free-energy difference, in a manner that obeys
detailed balance:
\begin{equation}
k_{s_j\to s^\prime_j}(\theta_j)=k_0w(\theta_j)\exp[\Delta U_{s_js_j^\prime}(\theta_j)/2], \,\,\,s_j,
s_j^\prime = 0, 1
\end{equation}
Here, $k_0$ sets the basic time scale, and $\Delta U_{s_js_j^\prime} ( \theta_j)
= U_{s_j^\prime}( \theta_j) - U_{s_j}( \theta_j)$. The function $w(\theta_j)$
describes the proton hopping windows (Fig. \ref{fig:OsterModel}), which reflect the idea that the proton
channel through the stator is gated by the motion of the rotor \cite{Xing:2006dq,Kojima01}.

Fig. S1 of the {\em Supporting Information} shows that this model,
developed by Oster and coworkers \cite{Xing:2006dq}, accurately describes the
torque-speed relation of the flagellar motor of {\em E. coli}.

\section{The rotor switching dynamics}
In {\em E. coli}, the fraction of time the motor rotates in the
clockwise direction, the so-called clockwise (CW) bias, is controlled
by the concentration of the intracellular messenger protein \Yp. This
protein modulates the CW bias by binding to the ring of
FliM proteins. This ring is connected to the ring of FliG proteins,
which interact with the stator proteins (Fig. \ref{fig:cartoon}).

The molecular mechanism of the switch is unknown. Yet, it is
widely believed that the binding of \Yp\ to FliM
tends to change the conformation of FliM. Following earlier
work, we assume that each FliM protein can exist in either a CW or CCW
conformational state and that binding of \Yp\ shifts the relative
stability of these two conformational states
\cite{Scharf98,Turner99,Duke01}. Moreover, we assume that also each
FliG protein can exist in either a CW or CCW conformational state. 
In the spirit of an MWC model \cite{Monod65}, we assume that  
 the energetic cost of having two rotor 
protein molecules in two different conformational states is
prohibitively large. 
We can then speak of the rotor being in either the CW or the CCW state.

When the rotor complex switches from one state to another, the
interactions between the FliG proteins and the stator proteins change,
due to the new conformational state of the FliG proteins. In our
model, each stator-rotor interaction is described by 4 energy
surfaces, $U_{s_j}^{r}$, with the subscript $s_j=0, 1$ denoting the
conformational state of the stator protein $j$ and the superscript
$r=0, 1$ denoting the conformational state of the rotor protein
(clockwise or counter-clockwise). We assume that the two rotor
surfaces corresponding to a given state of the stator are simply each
other's mirror image (the potentials are flipped in the $\theta$
direction), but offset by an energy difference $\Delta U_{\rm bias}$
that is given by the CW bias $P_{\rm CW} = \exp(-\beta \Delta U_{\rm
  bias})/(1+\exp(-\beta \Delta U_{\rm bias})$
(Fig. \ref{fig:OsterModel}B). This yields the following instantaneous rotor
switching rate:
\begin{equation}
\label{eq:k_s_r}
k^{r\to r^\prime} (\{\theta_j\})=\tilde{k}_0\exp[\Delta U^{rr^\prime}(\{\theta_j\})/2],
\,\,r, r^\prime = 0,1,
\end{equation}
where $\Delta U^{rr^\prime}(\{\theta_j\})=\sum_{j=1}^{N_{\rm S}} U_{s_j}^{r^\prime} (\theta_j)
-U_{s_j}^r(\theta_j)$. Importantly, the instantaneous switching rate
does not depend on the load, although the average, effective
switching rate does, as discussed
below.  Note also that our model assumes that the kinetics of ${\rm
  CheY_p}$ binding is fast on the time scale of switching, in contrast to a
recent model \cite{Tu08}.

\begin{figure}[t]
{\bf \sf A}\includegraphics[width=4cm]{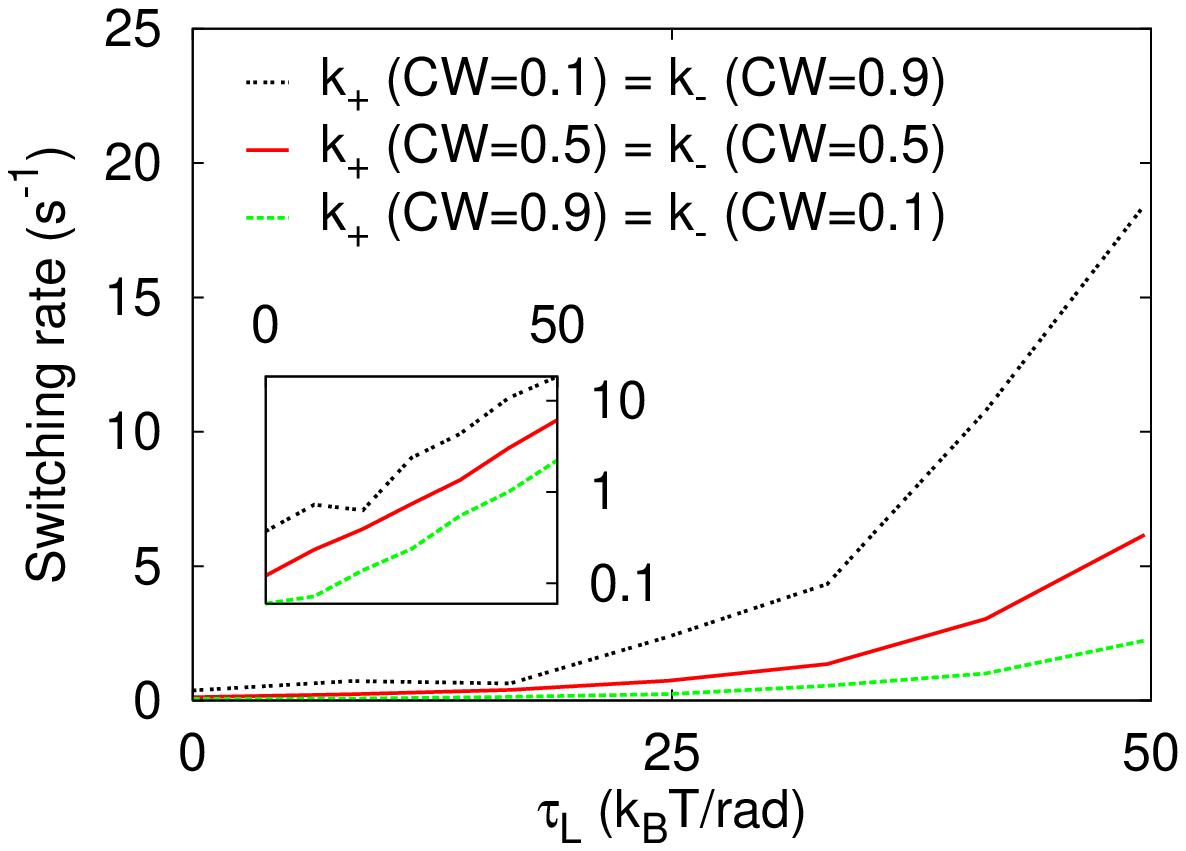}
{\bf \sf B}\includegraphics[width=4cm]{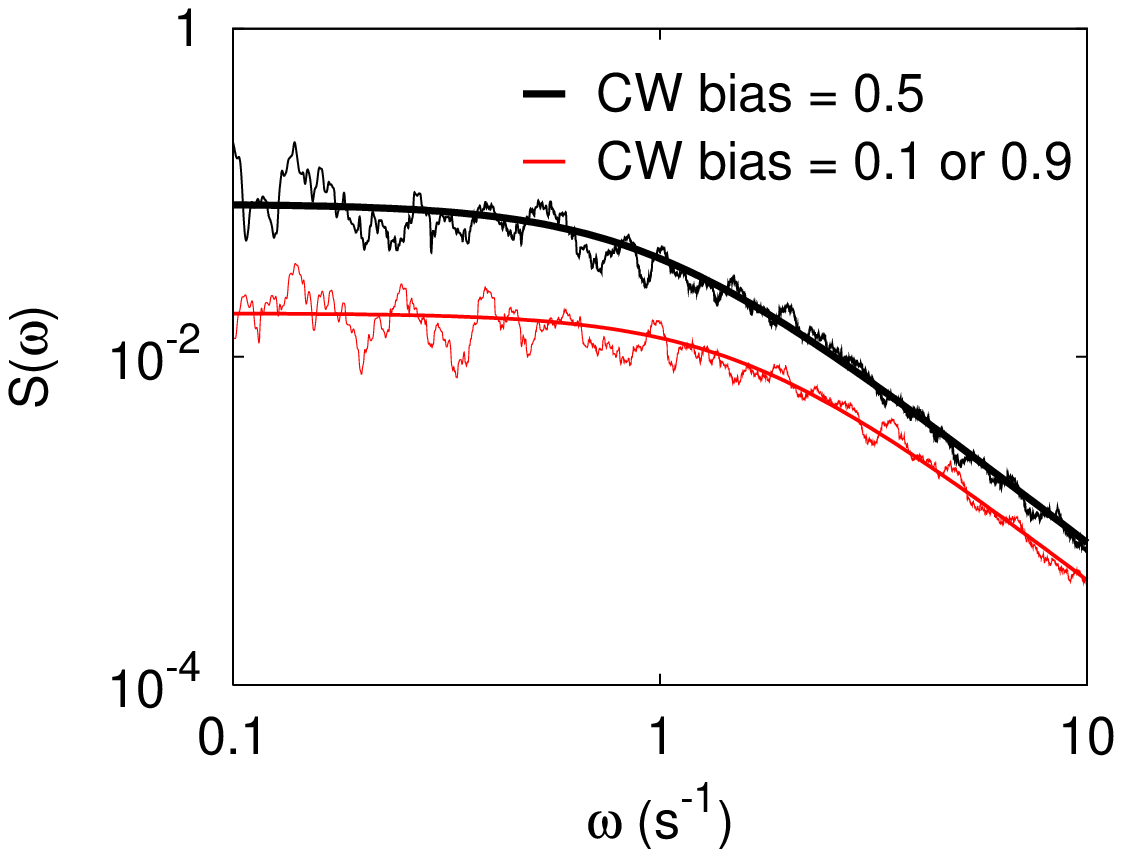}
\caption{\label{fig:switch_cons} Switching dynamics with conservative
  load - the load is constant in magnitude, but instantaneously
  changes sign upon a rotation reversal. A: Switching rate as a
  function of the load $\tau_{\rm L}$ in the
  forward CW $\to$ CCW ($k_+$) and backward
  (CCW $\to$ CW) direction ($k_-$) for CW bias = 0.1, 0.5, and 0.9. Note that
  due to the symmetry of our model, the switching dynamics in the
  forward (backward) direction for CW bias = $x$, equals the switching
  dynamics in the backward (forward) direction for CW bias = $1-x$. B:
  Power spectra $S(\omega)$ for CW bias = 0.1, 0.5 and 0.9.}
\end{figure}

Fig. \ref{fig:switch_cons} shows the switching dynamics when the load
is conservative. The conservative load is modeled as a constant torque
in a direction opposite to that of the rotation direction of the rotor
(Fig. \ref{eq:Lan}); after the rotor has switched direction, the
conservative force {\em instantaneously} changes sign. This means
that the switching dynamics of the motor with a conservative load
reflects the switching dynamics of the rotor complex; in the next
section, we study the effect of the flagellum, by studying the
switching dynamics using a viscous load, which depends on the
dynamics of the flagellum.

Fig. \ref{fig:switch_cons}A shows the average switching rate in the
forward and backward switching direction as a function of the
conservative load. As expected, the CCW $\to$ CW switching rate
increases as the CW bias increases. More interestingly, the switching
rate increases exponentially with the external load. As we describe below, this
is key to understanding the bump in the power spectrum.

The exponential dependence of the switching rate on the load can be
understood by noting that the effective switching rate is given by
\begin{equation}
\label{eq:k_switch}
k_{\rm switch}^{r \to r^\prime} = \int d\theta_{\rm R} P (\theta_{\rm R}) k^{r\to
  r^\prime}(\{\theta_j\}),
\end{equation}
where $P (\theta_{\rm R})$ is the stationary distribution of the
rotor's position. Increasing the load shifts $P (\theta_{\rm R})$ to
values of $\theta_{\rm R}$ where the driving force for switching,
$\Delta U^{r \to r^\prime}$, is larger (Fig. \ref{fig:OsterModel}B).
Since the instantaneous switching rate $k^{r\to
  r^\prime}(\{\theta_j\})$ depends exponentially on $\Delta U^{r \to
  r^\prime}(\{\theta_j\})$ (See Eq. \ref{eq:k_s_r}), the effective
switching rate $k_{\rm switch}^{r\to r^\prime}$ increases strongly
with load in the low-load regime, as observed in the experiments of Fahrner
{\em et al.} \cite {Fahrner03}.

These experiments also suggest that in the high torque regime, the
switching rate decreases with the load \cite{Fahrner03}, which is not
captured by the model presented here; yet, we believe that if we would
introduce a switching window analogous to the proton hopping window
introduced by Xing {\em et al.} to describe the torque-speed relation
\cite{Xing:2006dq}, we could reproduce this. However, this is not
critical for the problem considered here, since the switching
dynamics of a motor with a flagellum, as described in the next
section, is mostly determined by how the switching rate changes with
load in the low to intermediate torque regime, shown in
Fig. \ref{fig:switch_cons}.

Fig. \ref{fig:switch_cons}B shows the power spectra of the switching
dynamics. It is given by a Lorentzian, which shows that the switching
of the rotor without a flagellum can be modeled as a random telegraph process. 

\section{Flagellum dynamics}
In the model discussed above, after a
switching event, the torque on the motor immediately changes sign and
instantaneously reaches its steady-state value. However, in the
experiments of Cluzel and coworkers, the switching of the motor was
visualized via a bead that was attached
to the flagellar filament \cite{Korobkova04,Korobkova06}. We argue that the
flagellum dynamics is critical for understanding the switching
dynamics of the flagellar motor. 

Darnton and Berg recently studied polymorphic transitions of a single
 filament using optical tweezers \cite{Darnton:2007hc}. The
following three observations were made: 1) The transitions occur in
discrete, rapid steps that are stochastic in nature, suggesting that
they are activated processes during which an energy barrier is
crossed; 2) In between the steps, the filament behaves as a linear
elastic object that accumulates elastic strain energy that is released
during the next transformation; 3) During a step, not the whole
filament is converted, but micrometer-long sections.

\begin{figure}[t]
\includegraphics[width=6cm]{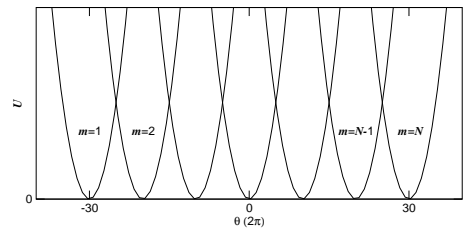}
\caption{\label{fig:FlagModel} Energy surfaces for the flagellum. The
  left-most curve ($m=1$) corresponds to the normal state, the
  right-most curve ($m=N$) corresponds to the curly state, while the
  intermediate conformational states correspond not only to the
  semi-coiled state, but also to hybrid filaments consisting of
  different sections of these polymorphic forms, as observed in the
  pulling experiment of Darnton and Berg \cite{Darnton:2007hc}. The
  polymorphic transitions are modeled as stochastic jumps between
  these energy surfaces (see Eq. \ref{eq:k_mm'}).}
\end{figure}

On the basis of these three observations, we have constructed the filament
model shown in Fig. \ref{fig:FlagModel}. It consists of a
number of harmonic potentials as a function of the winding angle
$\theta$, corresponding to different conformational states of the
filament. The left-most well corresponds to the normal state, which
is the polymorphic form of the filament when the motor runs in the
CCW direction. The right-most well corresponds to the curly state,
which is one of the polymorphic forms that the filament adopts when
the motor runs in the CW direction. The states in between correspond
to an ensemble of polymorphic forms that includes not only the coiled
and semi-coiled states, but, according to observation no. 3 above,
also states in which one section of the filament is in one distinct
polymorphic state, while the other is in an another
conformational state. According to observation no. 2, and following
Darnton and Berg \cite{Darnton:2007hc}, we assume that the free energy
$U^{\rm F}$ of a filament in a given polymorphic state $m$ is
quadratic in the curvature and torsion (see {\em Supporting
  Information}). The curvature $\kappa$ and torsion $\tau$ are
functions of the height of the filament $z$ and the winding angle
$\theta$. We assume that at each instant, the height
has relaxed to its steady-state value, which means that $U^{\rm
  F}$ becomes a quadratic function of the winding angle only:
\begin{equation}
U^{\rm F}_m(\theta) = \frac{1}{2} k_\theta (\theta - \theta_m)^2,
\end{equation}
where the torque constant $k_\theta$ is given by the Young's and shear
moduli and the contour length of the filament; the value chosen is
consistent with the measurements of Block {\em et al} \cite{Block:1989fe}
and Darnton and Berg \cite{Darnton:2007hc} (see {\em Supporting
  Information}). For simplicity, we assume that the potentials are
equally spaced, and have the same torque constant and well depth,
although under neutral pH the normal state is the most
stable one \cite{Darnton:2007hc}. The
total difference in winding angle between the normal (left-most)
and curly (right-most) state is about 50 rounds, which is the correct order
of magnitude based on the elastic properties of the filament (see
{\em Supporting Information}). This is an important parameter, since
it directly affects the characteristic switching time.

Motivated by observation 1, we assume that the transition from one
conformational state to another is an activated process, with a rate
constant
\begin{equation}
k_{m\to m^\prime}(\theta) = \breve{k}_0 \exp[(U^{\rm
  F}_m(\theta)-U^{\rm F}_{m^\prime}(\theta))/2].
\label{eq:k_mm'}
\end{equation}

Denoting the position of the load (bead)
with $\theta_{\rm L}$, the load dynamics is given by
 \begin{equation}
\gamma_{\rm L} \frac{d\theta_{\rm L}}{dt}= -k_\theta
(\theta_{\rm L} - \theta_{\rm R} -\theta_{m}) + \eta_{\rm L} (t).
\label{eq:viscous_load}
\end{equation}
Here, $\gamma_{\rm L}$ is the friction coefficient of the bead, and
$\eta_{\rm L}$ is a Gaussian white noise term of magnitude $\sqrt{2
  k_{\rm B}T \gamma_{\rm L}}$. 

\begin{figure}[t]
{\bf \sf A}\includegraphics[width=4cm]{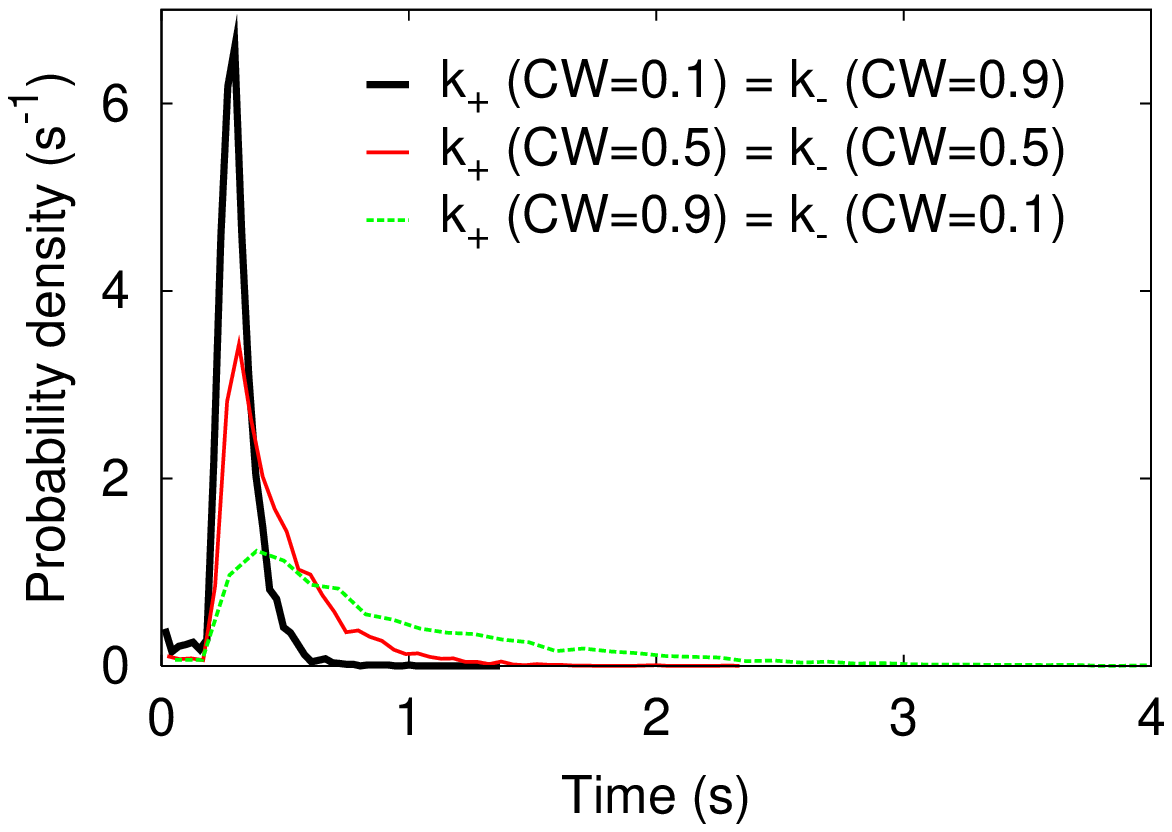}
{\bf \sf B}\includegraphics[width=4cm]{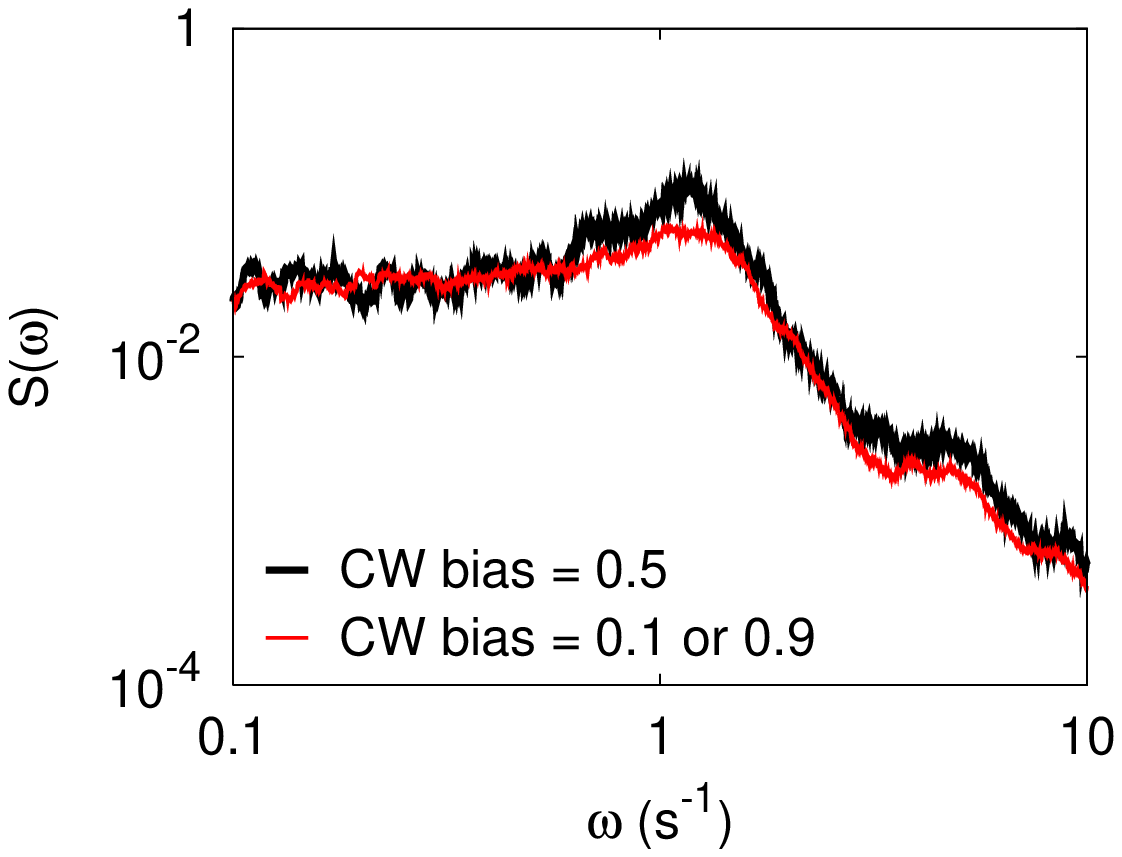}
\caption{\label{fig:switch_flag} The switching dynamics of a motor
  with viscous load (see Eq. \ref{eq:viscous_load}). Panel A:
  Distribution of waiting times for the forward CW $\to$ CCW
  transition ($k_+$) and backward CCW $\to$ CW transition ($k_-$), for
  CW bias = 0.1, 0.5 and 0.9. Panel B: The power spectra $S(\omega)$
  for CW bias = 0.1, 0.5 and 0.9.  Our model is symmetric by
  construction---the CW energy surface is the mirror image of the CCW
  surface (Fig. \ref{fig:OsterModel}B) and the wells of the
  filament potential are of equal depth (Fig. \ref{fig:FlagModel}). Accordingly,
  the distribution of the forward
  (backward) transition for CW bias = $x$ overlaps with that of the
  backward (forward) transition for CW bias = $1-x$. }
\end{figure}

Fig. \ref{fig:switch_flag} shows the switching characteristics of this
system. We show the dynamics of the bead instead of the motor, since
that has been measured experimentally; however, the switching dynamics
of the two are very similar.  Fig. \ref{fig:switch_flag}A shows the
distribution of waiting times, for CW bias = 0.1, 0.5, and 0.9. These
distributions agree remarkably well with those observed by Cluzel and coworkers  \cite{Korobkova06}. Firstly, the distributions are not
exponential, as would be expected for a random telegraph process:
The
distributions exhibit a clear peak at around $0.4~{\rm
  s}^{-1}$. Secondly, the waiting-time
distribution for the forward (CW$\to$CCW) transition shifts from a
narrow distribution at CW bias = 0.1 to a broad distribution at CW
bias = 0.9. Moreover, the position of the maximum of the distribution
shifts to longer times. All these features are in near quantitative
agreement with experiment. 

Fig. \ref{fig:switch_flag}B shows the power spectra of our model, for
CW bias = 0.5, and for CW bias = 0.1, 0.9 (they are identical because
of the symmetry of our model). It is seen that the spectra
exhibit a distinct peak at $\omega \sim 1~{\rm s}^{-1}$. Moreover, the
peak is most pronounced when the CW bias = 0.5. These characteristics
are observed experimentally \cite{Korobkova06}.

\begin{figure}[t]
{\bf \sf A}\includegraphics[width=4cm]{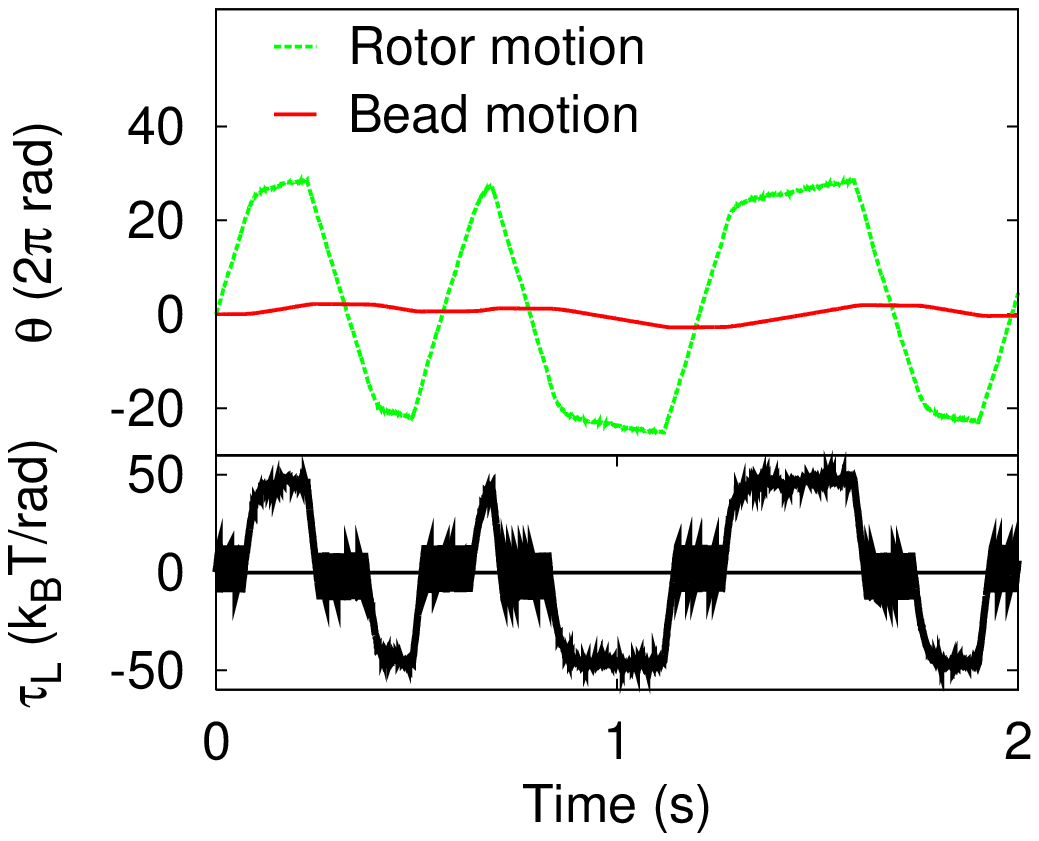}
{\bf \sf B}\includegraphics[width=4cm]{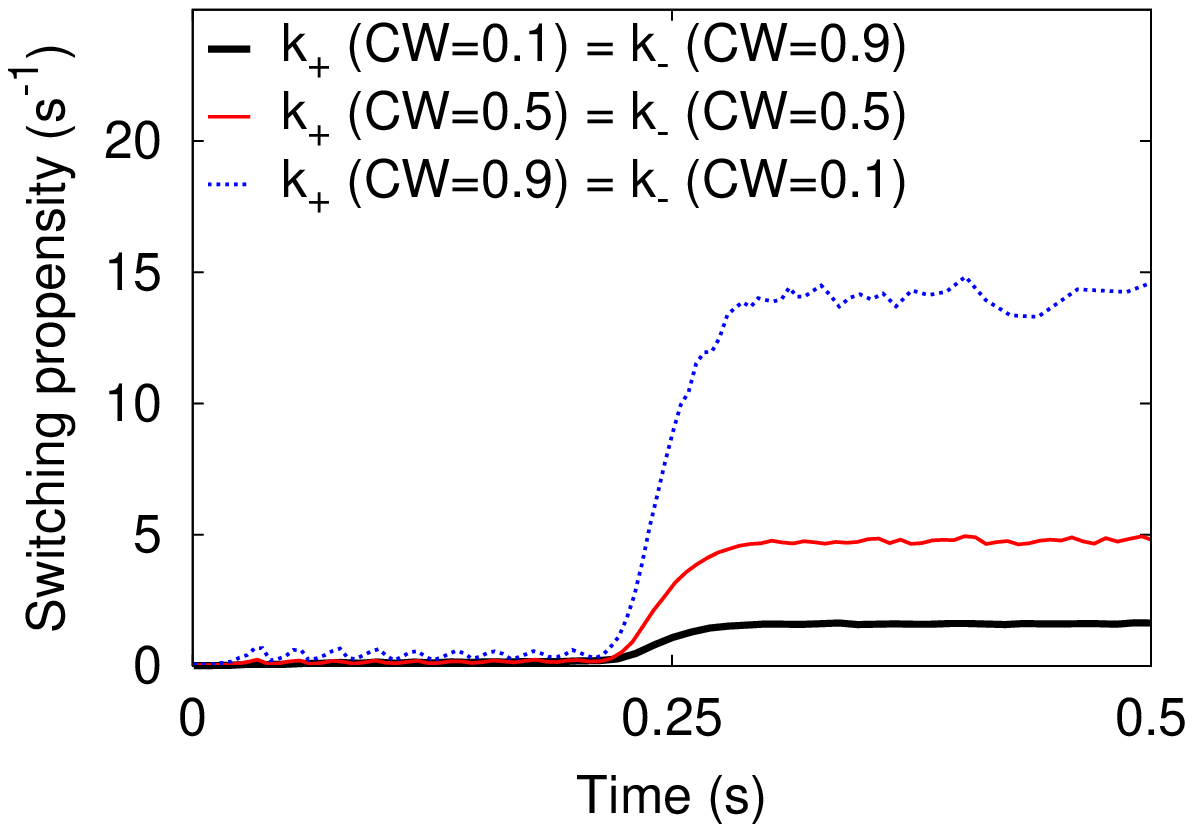}
\caption{\label{fig:explanation} The mechanism of switching. A:
  Typical time traces for the torque, motor and bead position, for CW
  bias = 0.5.  B: The switching propensity as a function of time
  after a switching event, for CW bias = 0.1, 0.5, and 0.9. This
  figure is obtained by combining the switching propensity as a
  function of the conservative load, as shown in
  Fig. \ref{fig:switch_cons}A, with the average force as a function of time
  after a switching event, as shown in panel A.}
\end{figure}

Our model predicts that the peak in the power spectrum arises from the
interplay between the conformational dynamics of the flagellum and the
dependence of the switching rate on the load
(Fig. \ref{fig:switch_cons}A). The idea is illustrated in
Fig. \ref{fig:explanation}.
After a switching event of the rotor, the torque is initially in the
original direction, but decreases rapidly in magnitude as the
filament reaches its optimum winding angle (Fig. \ref{fig:FlagModel};
Fig. \ref{fig:explanation}A); in this regime, the load
on the motor is negative, and the elastic strain energy in the
filament decreases.  As the rotor drives the filament beyond its
optimal winding angle, the torque changes direction and increases in
magnitude; the load on the motor becomes positive, and the strain
energy in the filament builds up. This strain energy can, however, be
released via a polymorphic transition, leading to a sudden change in
the direction of the torque. This process repeats itself until the
filament reaches its final polymorphic form, upon which the strain
energy can no longer be released, and the torque increases to reach a
plateau when the viscous drag equals the motor torque. The peak in the
power spectrum can now be understood by combining the time trace of
the torque (Fig. \ref{fig:explanation}A) with the dependence of the
switching rate on the load (Fig. \ref{fig:switch_cons}A), yielding the
switching propensity ({\em i.e.}, the probability to switch per unit
amount of time at a given time $t$) as a function of time, as shown in
Fig. \ref{fig:explanation}B. After a switching event, the torque
flip-flops around zero and the switching propensity remains low. But
when the flagellum has reached its final polymorphic form, the torque
can no longer be released, and the switching propensity increases
significantly. The peak in the spectrum is precisely caused by the
fact that the switching propensity function is not constant in time,
as for a Markovian Poisson process, but increases with time.

\section{Coarse-grained model}
Our calculations suggest that a useful coarse-grained model for
understanding the
 switching dynamics is one in which the system stochastically
flips between two states with time-dependent propensity functions
(Fig. \ref{fig:explanation}B):
\begin{equation}
{\rm CW} \overset{k_+(t)}{\underset{k_-(t)}
  {\rightleftharpoons}} {\rm CCW},
\label{eq:cg}
\end{equation}
where the propensity functions are given by the following piece-wise
linear functions:
\begin{align}
k_{\alpha}(t) &= k_\alpha^{\rm min} &  t < t_1\\
k_{\alpha}(t)&= k_\alpha^{\rm min} + (k_{\alpha}^{\rm
  max}-k_{\alpha}^{\rm min}) \frac{t-t_1}{t_2-t_1} &t_1 < t < t_2\\
k_{\alpha}(t) &= k_\alpha^{\rm max} &  t > t_2
\label{eq:cg_k}
\end{align}
The important parameters of this model are the lag time,
$T_{\alpha}=(t_1+t_2)/2$, the minimum and maximum propensity,
$k_{\alpha}^{\rm min}$ and $k_\alpha^{\rm max}$, respectively, and to a lesser extent the sharpness of
the transition $s_{\alpha}=(k_{\alpha}^{\rm max}-k_{\alpha}^{\rm
  min})/(t_2-t_1)$. For this model,  the
waiting-time distribution and power spectrum can be obtained
analytically (see {\em Supporting
  Information}).

The maximum propensity function $k_\alpha^{\rm max}$ depends on the
CW bias, which determines how the rotor switching propensity depends
upon the load (Fig. \ref{fig:switch_cons}A), and the maximum load
itself, which is determined by the drag coefficient of the load,
$\gamma_{\rm L}$, and the torque-speed relation; to a good
approximation, the maximum torque is given by the intersection of
$\gamma_{\rm L}$ times the speed (the load line) and the torque-speed
curve \cite{Xing:2006dq}. The minimum propensity function
$k_\alpha^{\rm min}$ does not
depend upon this maximum load, but rather upon the torque at which the
flagellum undergoes a polymorphic transition---the polymorphic
transitions release the elastic strain energy before this maximum
load is reached, keeping the load and hence $k_\alpha^{\rm min}$
low.
The peak in the power spectrum arises when $k_\alpha^{\rm min} <
k_\alpha^{\rm max}$, which leads to the prediction that the peak may
disappear when the number of stators is reduced. The position of the
peak is determined by $T_\alpha$, which is given by the difference in
winding angle between the normal and curly state divided by the
average speed at which the rotor drives the systems between these two
states; this speed depends upon the torque-speed relation and the load
as a function of time. Interestingly, polymorphic transitions of
filaments of swimming bacteria occur on time scales of 0.1 s
\cite{Turner00}, close to the peak of the waiting-time distribution,
supporting our idea that they set the characteristic switching time.

The dependence of $k_\alpha^{\rm max}$ on the CW bias can explain the
change in the waiting-time distributions when the CW bias is varied.
(Fig. \ref{fig:switch_flag}A). When the CW bias is large, $k_-^{\rm
  max}$ is large, because the switching propensity at the maximum load
as set by the balance of the drag and the motor torque is large
(Fig. \ref{fig:switch_cons}A). Consequently, the rotor typically
switches to the CW state before the switching propensity can reach
its plateau value. This explains the narrow distribution of CCW
intervals when the CW bias is large, as observed in both the model
(Fig. \ref{fig:switch_flag}A) and experiment \cite{Korobkova06}.  For
the reverse transition the situation is qualitatively different. When
the CW bias is large, $k_+^{\rm max}$ is low, which means that the
system can enter the 
regime in which the switching propensity is constant before it
switches to the CCW state. This constant
propensity leads to an exponential tail in the distribution of CW
(CCW) intervals when the CW (CCW) bias is large, as observed in both
the distributions of the model (Fig. \ref{fig:switch_flag}A) and those
measured experimentally \cite{Korobkova06}.

\section{Discussion}
We have presented a statistical-mechanical model that describes the
switching dynamics of a rotary flagellar motor. Its foundation is the
assumption that the rotor protein complex can exist in two
conformational states corresponding to the two respective rotation
directions, and that switching between these states depends on
interactions with the stator proteins, which also drive the rotation
of the rotor complex. This naturally couples the switching dynamics to
the rotation dynamics. The load does not directly change the relative
stability of the rotor's conformational states, but it does change how
often the stator proteins during their motor cycle favor one
conformational state of the rotor over the other. This, according to
our model, is the principal mechanism that makes the switch sensitive
to torque and speed. Another key element of our model is
that after a switch, it takes time for the load to build up, due to
the polymorphic transitions of the filament.

Several predictions emerge from our model that could be tested
experimentally. One is that the change in the torque on the filament
upon a motor reversal leads to a series of polymorphic transitions,
which could be tested by applying a torque on a single filament using
magnetic tweezers. Another is that the switching dynamics of the
rotor without the viscous load of the flagellum is that of a two-state Poisson process, in contrast to a
recent model \cite{Tu08}; moreover, 
 in the zero-load regime, the switching propensity increases with
the load. These predictions could be tested by measuring the rotation
dynamics of a bead that is connected either directly to the stub, or
to a very short filament. Our model predicts that its power spectrum
does not have a peak.

We thank Howard Berg, Sanne Abeln and Gijsje Koenderink for a critical
reading of the manuscript in this project. The work is supported by
FOM/NWO.

\bibliographystyle{pnas}
\bibliography{/Users/tenwolde/articles/references/sysbio}

\end{document}